 \documentclass[final,3p,times,twocolumn]{elsarticle}
\usepackage{amssymb}
\usepackage{multirow}
\usepackage{amsmath}
\usepackage{url}
\journal{Astroparticle Physics}

\begin{document}

\begin{frontmatter}



\title{Antideuteron Sensitivity for the GAPS Experiment}


%\iffalse
\author[CAL]{T. Aramaki\corref{SLAC}} 
%\fnref{}
\cortext[SLAC]{\textit{Current institution: SLAC National Accelerator Laboratory/Kavli Institute for Particle Astrophysics and Cosmology, Menlo Park, CA 94025, USA}}
\ead{tsuguo@slac.stanford.edu}
\author[CAL]{C.J. Hailey}
\author[UCB]{S.E. Boggs} 
%\author[UCB]{W. W. Craig}
\author[Hawaii]{P. von Doetinchem}
%\author[ORNL]{L. Fabris} 
\author[JAXA]{H. Fuke} 
%\author[Latvia]{F. Gahbauer}
%\author[CAL]{J.E. Koglin}
%\author[CAL]{N. Madden}
\author[UCLA]{S.I. Mognet}
%\author[CAL]{K. Mori}
\author[UCLA]{R.A. Ong}
\author[CAL]{K. Perez}
%\author[JAXA]{T. Yoshida}
%\author[UCLA]{T. Zhang}
\author[UCLA]{J. Zweerink} 
%\author[CAL]{H. T. Yu}
%\author[ORNL]{K. P. Ziock}

\address[CAL]{Columbia Astrophysics Laboratory, Columbia University, New York, NY 10027, USA}
%\address[SLAC]{SLAC National Accelerator Laboratory/Kavli Institute for Particle Astrophysics and Cosmology, Menlo Park, CA 94025, USA}
\address[UCB]{Space Sciences Laboratory, University of California, Berkeley, CA 94720, USA}
\address[Hawaii]{Department of Physics and Astronomy, University of Hawaii at Manoa, Honolulu, HI 96822 USA}
\address[JAXA]{Institute of Space and Astronautical Science, Japan Aerospace Exploration Agency (ISAS/JAXA), Sagamihara, Kanagawa 252-5210, Japan}
\address[UCLA]{Department of Physics and Astronomy, University of California, Los Angeles, CA 90095, USA}

%\address[LLNL]{Lawrence Livermore National Laboratory, Livermore, CA, USA}
%\address[Latvia]{Department of Physics, University of Latvia, Riga, LV 1586, Latvia}
%\address[ORNL]{Oak Ridge National Laboratory, Oak Ridge, TN 37831, USA}

%\fi

\begin{abstract}
The General Antiparticle Spectrometer (GAPS) is a novel approach for indirect dark matter searches that exploits cosmic antiparticles, especially antideuterons. The GAPS antideuteron measurement utilizes distinctive detection methods using atomic X-rays and charged particles from the decay of exotic atoms as well as the timing and stopping range of the incoming particle, which together provide excellent antideuteron identification. Prior to the future balloon experiment, an accelerator test and a prototype flight were successfully conducted in 2005 and 2012 respectively, in order to verify the GAPS detection concept. This paper describes how the sensitivity of GAPS to antideuterons was estimated using a Monte Carlo simulation along with the atomic cascade model and the Intra-Nuclear Cascade model. The sensitivity for the GAPS antideuteron search obtained using this method is 2.0 $\times 10^{-6}$ [m$^{-2}$s$^{-1}$sr$^{-1}$(GeV/$n$)$^{-1}$] for the proposed long duration balloon program (LDB, 35 days $\times$ 3 flights), indicating that GAPS has a strong potential to probe a wide variety of dark matter annihilation and decay models through antideuteron measurements. GAPS is proposed to fly from Antarctica in the austral summer of 2019-2020.

\end{abstract}


\end{frontmatter}

% \linenumbers

\section{Introduction}
\label{section: Introduction}

\subsection{Dark Matter and WIMPs}
\label{subsection: Dark Matter and WIMPs}

The recent result by the Planck experiment \cite{Planck2013} shows that 68\% of our universe is composed of dark energy, 27\% is dark matter and 5\% is baryonic matter. The nature and origin of dark energy and dark matter, however, are still unknown. This is one of the great cosmological problems of the 21st century. The existence of dark matter was postulated by Fritz Zwicky in 1933 from the observation of the Coma galaxy cluster \cite{Zwicky1937}. Recent observations, such as gravitational lensing in the Bullet Cluster (two colliding clusters of galaxies), also indicate  the existence of dark matter \cite{Clowe2004}. Weakly Interacting Massive Particles (WIMPs) are among the theoretically best-motivated candidates in the variety of dark matter models. The lightest supersymmetric partner (LSP) in supersymmetric (SUSY) theories such as neutralinos \cite{Donato2000,Donato2008} and right-handed sneutrinos \cite{Cerdeno2009,Cerdeno2014} as well as Kaluza-Klein particles, such as right-handed neutrinos (LZP) \cite{Baer2005} in extra dimension theories are examples of popular WIMP candidates. Gravitinos, known as Super Weakly Interacting Massive Particles (SuperWIMPs), are also favored dark matter candidates in SUSY theories \cite{Dal2014}.   

\subsection{Antideuterons as Dark Matter Search}
\label{subsection: Antideuterons for Dark Matter Search}

Dark matter searches are usually categorized into three types: particle colliders, direct searches, and indirect searches. The direct and indirect searches will measure the relic WIMPs in our universe, while the particle collider will try to create WIMPs. The detection methods and the background models for each search are different, but also complementary, helping to illuminate the nature of dark matter. 

Indirect dark matter searches measure dark matter co-annihilation and decay products, such as electrons, positrons, gamma-rays, antiprotons and antideuterons. The General Antiparticle Spectrometer (GAPS) is a novel approach for indirect dark matter searches that exploits cosmic antiprotons and antideuterons \cite{Mori2002,Aramaki2014} .  

\begin{figure}[tbp]
\begin{center} 
\includegraphics*[width=7.5cm]{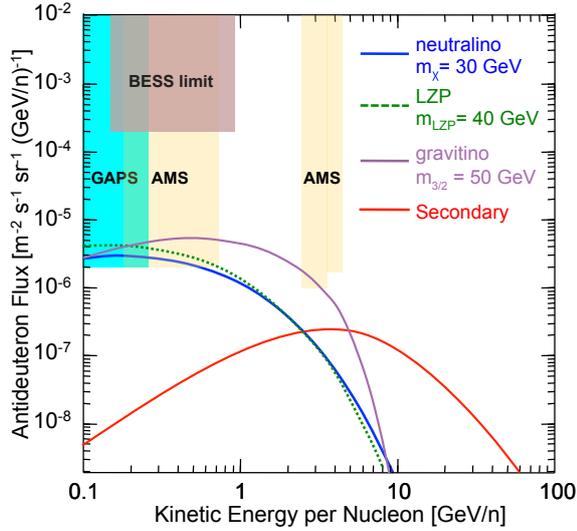}
\end{center}
\caption{Antideuteron flux at the top of the atmosphere. The solid blue and purple lines are neutralino and gravitino LSPs with $m_{\chi} \sim$ 30 GeV \cite{Donato2008} and $m_{3/2} \sim$ 50 GeV \cite{Dal2014} respectively, while the dashed green line is for LZP with $m_{\chi}$ $\sim$ 40 GeV \cite{Baer2005}. The red solid line represents the secondary/tertiary flux due to the cosmic-ray interactions \cite{Duperray2005,Salati2010,Ibarra2013}. The expected GAPS antideuteron sensitivity ($\sim$ 99\% CL) as well as the AMS-02 sensitivity \cite{Doetinchem2014,Mognet2014,Fuke2014,Aramaki2015} and the current upper limit on the antideuteron flux obtained by BESS \cite{Fuke2005} are also shown (see Section \ref{subsection: Sensitivity and Confidence Level}).}
\label{dbar_sensitivity}
\end{figure} 

Antideuteron production in dark matter co-annihilations was proposed by Donato et al., in 2000 \cite{Donato2000,Donato2008}. The antideuteron flux at the top of the atmosphere due to dark matter co-annihilation and decay (called primary flux) can be estimated based on dark matter density profiles in the galaxy, dark matter co-annihilation and decay channels, hadronization and coalescence models, propagation models and solar modulation models. The primary antideuteron fluxes for different dark matter models are shown in Figure \ref{dbar_sensitivity}. The solid blue and purple lines are neutralino and gravitino LSPs with $m_{\chi} \sim$ 30 GeV \cite{Donato2008} and $m_{3/2} \sim$ 50 GeV \cite{Dal2014} respectively, while the dashed green line is for LZP with $m_{\chi}$ $\sim$ 40 GeV \cite{Baer2005}. The primary antideuteron flux has a relatively flat peak at low energy,  $E \sim$ 0.2 GeV/$n$. The antideuteron flux due to cosmic-ray interactions with the interstellar medium (called secondary/tertiary flux) is also shown as the solid red line in Figure \ref{dbar_sensitivity} \cite{Duperray2005,Salati2010,Ibarra2013}. Unlike primary antideuterons, collision kinematics suppress the formation of low-energy secondary antideuterons. Furthermore, the interaction rate is drastically decreased since the flux of the cosmic-ray protons follows roughly a power law, $F_p \sim E^{-2.7}$. Therefore, the primary antideuteron flux can be about two orders of magnitude larger than the secondary/tertiary antideuteron flux at low energy, which can provide clear dark matter signatures. 

The Alpha Magnetic Spectrometer (AMS-02) on the International Space Station, launched in 2011, is the only currently-operating antideuteron search experiment. AMS-02 probes a higher energy region than GAPS, using a different detection technique with a completely different background \cite{Giovacchini2007}. GAPS complements AMS-02 on the antideuteron search, which is crucial for rare event searches, while uniquely seeking for lower energy and lower background antideuterons. GAPS also complements other existing and planned indirect searches as well as underground direct dark matter searches and collider experiments, while distinctively probing a wide variety of dark matter annihilation and decay models. The expected GAPS antideuteron sensitivity ($\sim$ 99\% CL) as well as the AMS-02 sensitivity \cite{Doetinchem2014,Mognet2014,Fuke2014,Aramaki2015} and the current upper limit on the antideuteron flux obtained by BESS \cite{Fuke2005} are also shown in Figure \ref{dbar_sensitivity}. The details will be discussed in Section \ref{subsection: Sensitivity and Confidence Level}.

\section{GAPS Project}
\label{section: GAPS Project}

The GAPS experiment was proposed in 2002 \cite{Mori2002}, and the detection concept discussed below was validated in an accelerator beam test at KEK, Japan in 2005 \cite{Aramaki2013} and the prototype flight (pGAPS) at JAXA/ISAS balloon facility in 2012 \cite{Doetinchem2014,Mognet2014,Fuke2014}. A GAPS science mission is proposed to fly from Antarctica in the austral summer of 2019-2020.

\subsection{Detection Concept}
\label{subsection: Detection Concept}

\begin{figure}[htbp]
\begin{center} 
\includegraphics*[width=7.5cm]{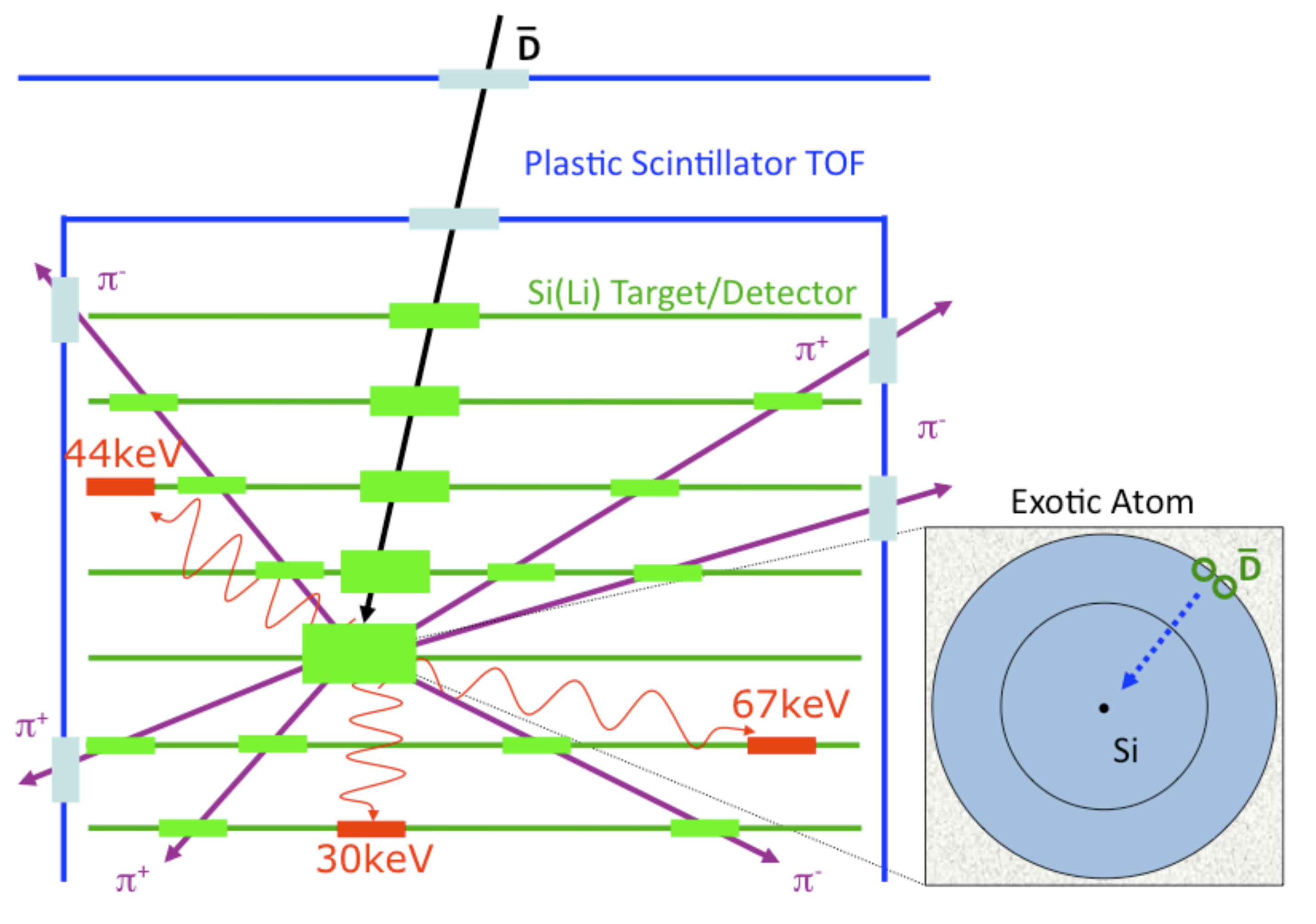}
\end{center}
\caption{GAPS detection method: An antiparticle slows down and stops in the Si(Li) target, forming an exotic atom. The atomic X-rays will be emitted as it de-excites followed by the pion (and proton) emission in the nuclear annihilation.}
\label{detection}
\end{figure}

The GAPS detection method involves capturing antiparticles into a target material with subsequent formation of exotic atoms. A time-of-flight (TOF) system measures the velocity (energy) and direction of an incoming antiparticle. The antiparticle slows down by the dE/dX energy loss and stops in the target material, forming an exotic atom in its excited state. The exotic atom de-excites with the emission of Auger electrons as well as atomic X-rays \cite{Aramaki2013}. With known atomic number of the target, the Bohr formula for the atomic X-ray energy uniquely determines the mass of the captured particle \cite{Mori2002}. Ultimately, the antiparticle is captured by the nucleus in the atom, where it is annihilated with the emission of annihilation products, such as pions and protons. Since the mean number of pions and protons produced by the nuclear annihilation is approximately proportional to the number of antinucleons, the pion and proton multiplicities provide an additional discriminant to identify the incoming antideutron from other particles, such as antiprotons. This process is illustrated in Figure \ref{detection}. 

\begin{figure}[htbp]
\begin{center} 
\includegraphics*[width=7.5cm]{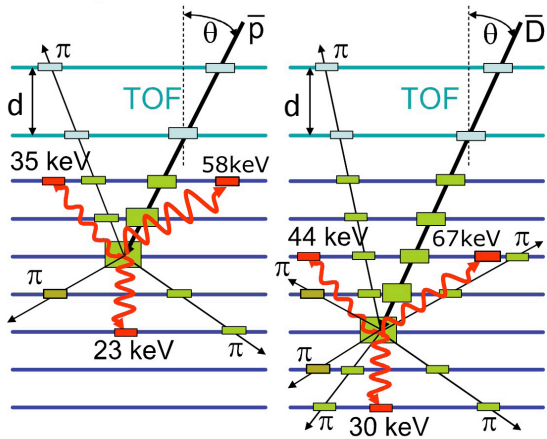}
\end{center}
\caption{GAPS Antideuteron-antiproton identification technique: (1) unique atomic X-rays, (2) pion and proton multiplicity, (3) stopping range. }
\label{identification}
\end{figure}

Antiprotons are the main background in the GAPS antideuteron measurement since they can also form exotic atoms and simultaneously emit atomic X-rays and annihilation products. Antideuterons, however, are distinguishable from antiprotons with the atomic X-rays and annihilation products as discussed above. Furthermore, the stopping range of antideuterons, which is about twice as large as that of antiprotons with the same $\beta$, provides an excellent discrimination to distinguish antideuterons from antiprotons. Figure \ref{identification} illustrates the GAPS antideuteron-antiproton identification technique with atomic X-rays, pion and proton multiplicities, and stopping range. The details will be discussed below.

\subsection{Instrumental Design}
\label{subsection: Instrumental Design}

\begin{figure}[htbp]
\begin{center} 
\includegraphics*[width=7.5cm]{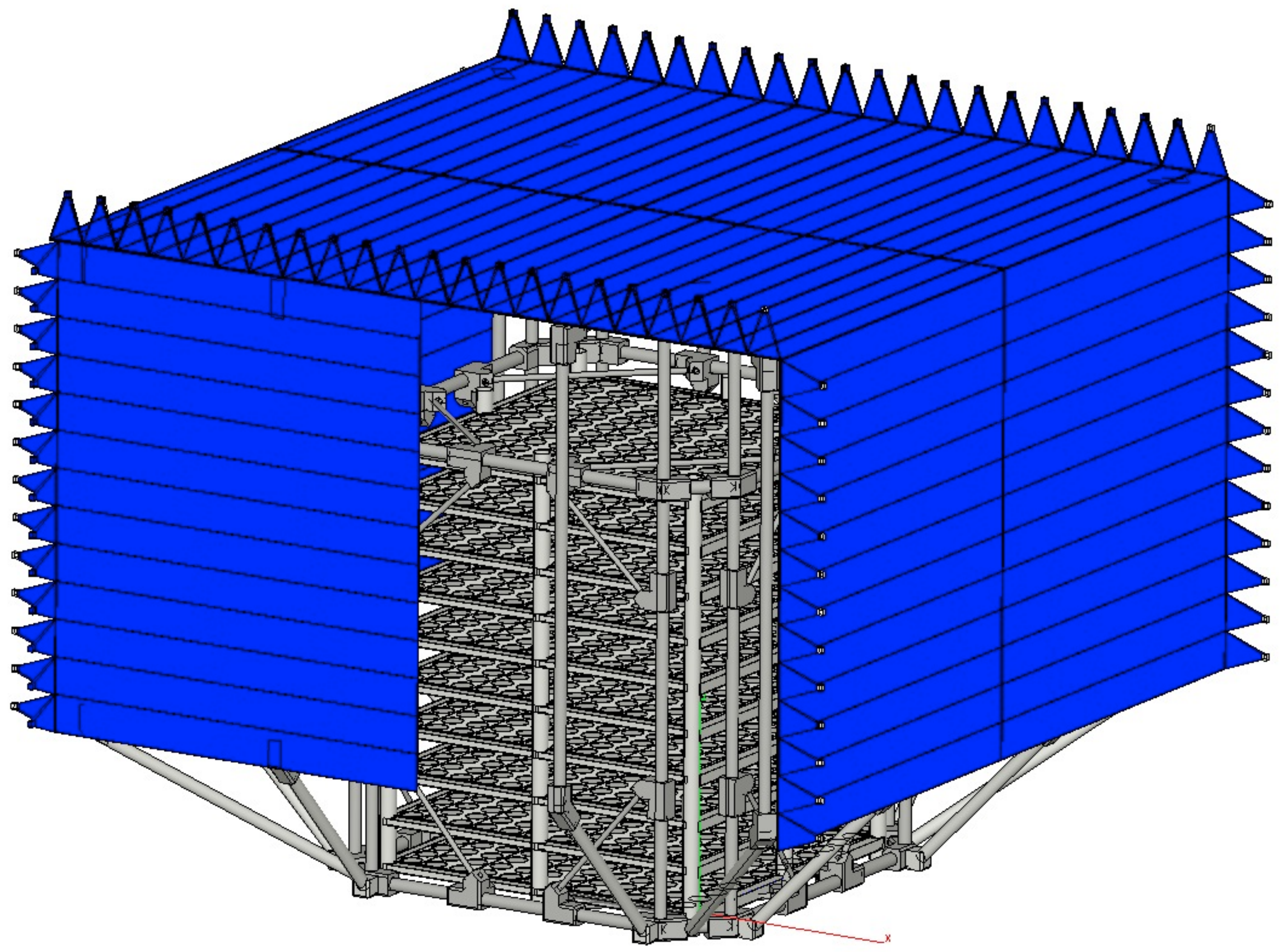}
\end{center}
\caption{GAPS detector design: 10 layers of Si(Li) detectors are surrounded by the inner and outer TOF plastic scintillators. Note that inner TOF paddles are not shown in the figure in order to show the detector layers. Each Si(Li) detector is 4 inch diameter, 2.5 mm thick.}
\label{detector}
\end{figure}

The GAPS balloon flight will have several unique features. It will be the first balloon flight with a very large, pixellated lithium-drifted silicon (Si(Li)) detector surrounded by a very large TOF system without a pressure vessel. The Si(Li) detector will be used both as a target to form the exotic atom as well as a detector for atomic X-rays and annihilation products. There will be 10 layers of detectors with each layer composed of 4 inch diameter, 2.5 mm thick Si(Li) detectors. Each Si(Li) detector will be segmented into 4 strips and adjacent tracking layers separated by 20 cm will have their strips positioned orthogonally. The inner TOF system (1.6 m $\times$ 1.6 m $\times$ 2 m) composed of highly segmented plastic scintillators will completely surround the Si(Li) layers. The top half of the inner TOF system and Si(Li) layers will be covered with plastic scintillators (outer TOF system, 3.6 m $\times$ 3.6 m $\times$ 2 m) and the inner and outer TOF planes will be separated by 1 m (see Figure \ref{detector}). This geometry is a natural consequence of the current GAPS detection concept, which allows us to track and count the number of particles produced in the nuclear annihilation, while separately identifying atomic X-rays from particle tracks. It also permits direct measurement of particle stopping range in the multi-layer geometry. 

\section{GAPS Antideuteron Sensitivity}
\label{section: GAPS Antideuteron Sensitivity}

As discussed above, antideuterons can be identified from other cosmic-ray particles by using the GAPS detection technique with atomic X-rays and pion/proton multiplicities, as well as the different stopping range of incoming particles. Low-energy cosmic-ray protons, even though their flux is several orders of magnitude larger than the antideuteron flux, can be clearly distinguishable from antideuterons since they cannot emit atomic X-rays nor annihilation products. Furthermore, the coincidence of multiple cosmic-ray protons coming into the same channel within a short time window ($\sim$ 100 ns) is negligible. Therefore, we will focus on antiprotons as the main background in the GAPS experiment as they can also form exotic atoms and produce atomic X-rays and charged particles. In the following section, we will discuss how antideuterons can be identified from antiprotons and also show the simple, but comprehensive approach to the GAPS antideuteron sensitivity estimation by using Monte Carlo simulations, atomic cascade model and Intra-Nuclear Cascade (INC) model. The GAPS antideuteron sensitivity was estimated with $\sim$ 99\% confidence level (CL) to identify primary antideuterons from other cosmic-ray particles. 

\subsection{Grasp}
\label{subsection: Grasp}

\begin{figure}[b]
\begin{center} 
\includegraphics*[width=7.5cm]{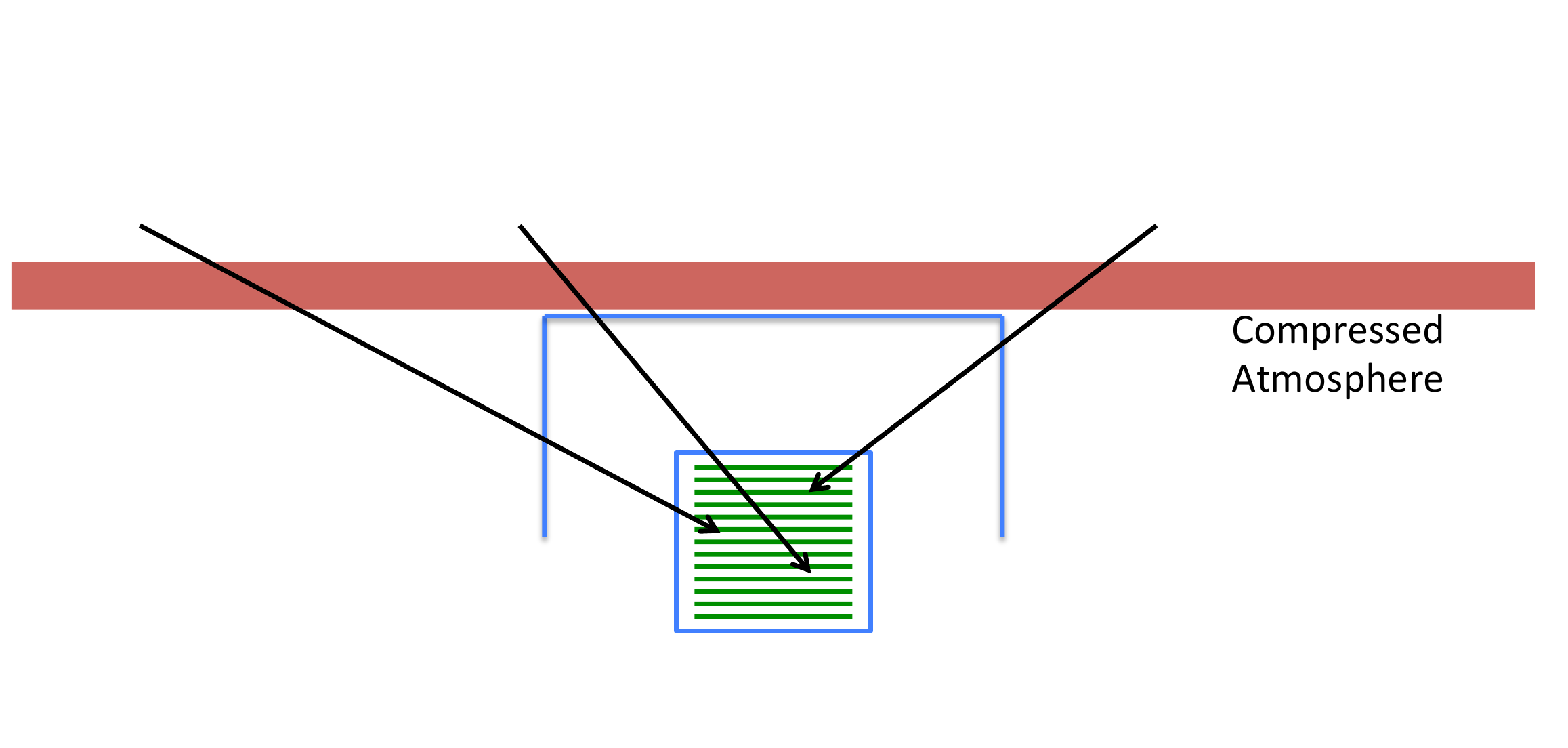}
\end{center}
\caption{The Geant4 setup for grasps estimation. Antiprotons and antideuterons were generated above the compressed atmosphere (a flat plane, column density $\sim$ 4g/cm$^2$) to evaluate the energy loss and in-flight annihilation in the atmosphere.}  
\label{Geant4}
\end{figure}

Antiparticles coming from the top of the atmosphere are degraded and also annihilated in-flight in the atmosphere before reaching the GAPS instrument. Some of them are able to reach and stop inside the GAPS instrument. A Geant4 Monte Carlo simulation was conducted to evaluate how many antiparticles can reach the GAPS instrument during the balloon flight. As seen in Figure \ref{Geant4}, antiprotons and antideuterons were generated above the compressed atmosphere (a flat plane, column density $\sim$ 4g/cm$^2$) to evaluate the energy loss and in-flight annihilation in the atmosphere. The grasps [$m^2 sr$] of the GAPS instrument at $\sim$ 40 km flight altitude for antiprotons and antideuterons that stop in the Si(Li) layer $\Gamma^{stop}$ and annihilate in-flight in the instrument (not in the atmosphere) $\Gamma^{inf}$ were estimated with Geant4.10.01. Note that $\Gamma^{stop}$ is defined as a product of the effective area, solid angle and stopping efficiency, while $\Gamma^{inf}$ is defined as a product of the effective area, solid angle and efficiency of the in-flight annihilation.

\begin{figure}[htbp]
\begin{center} 
\includegraphics*[width=7.5cm]{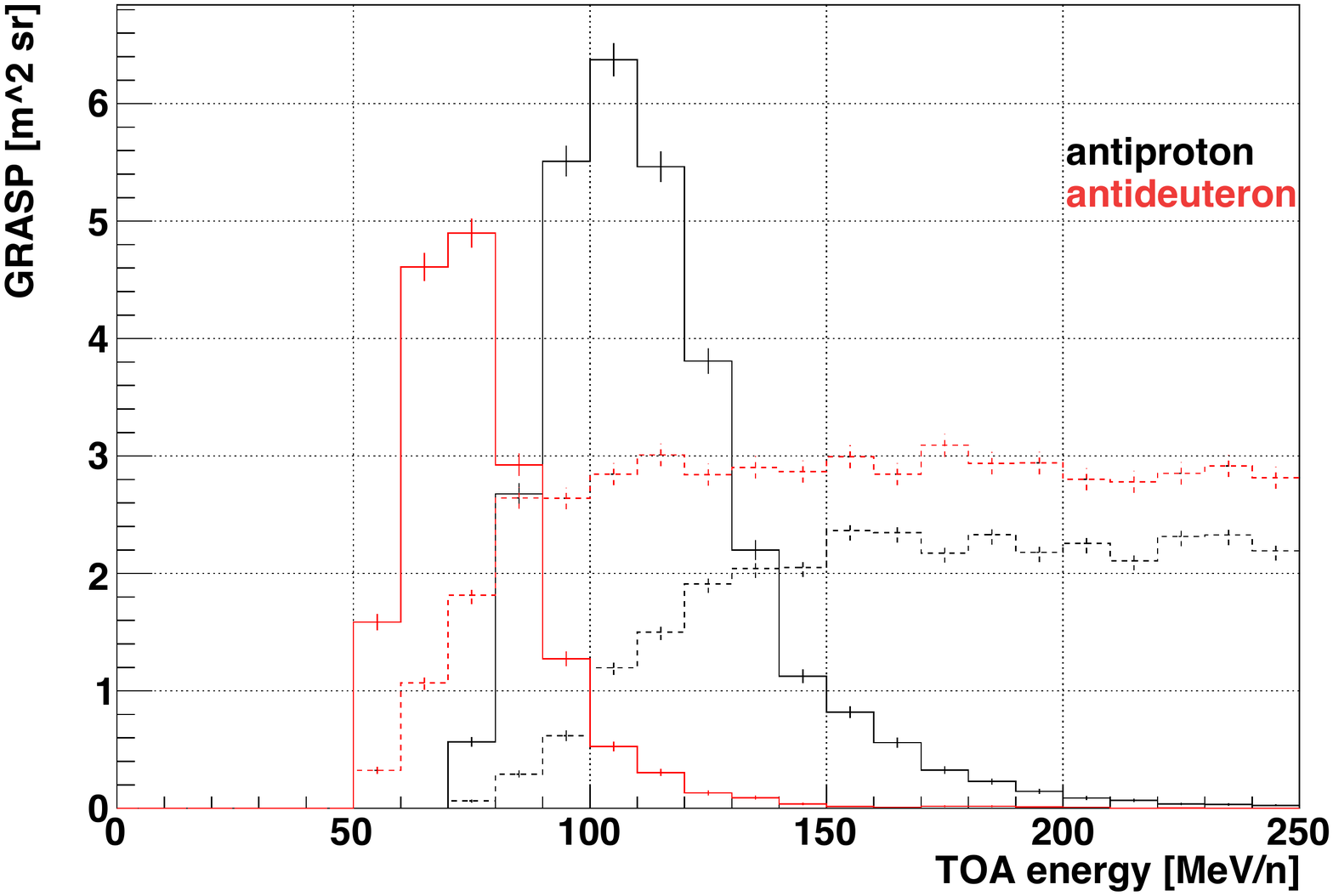}
\end{center}
\caption{The grasps for antiprotons (red lines) and antideuterons (black lines). The solid lines are grasps for stopped events, while the dashed lines are for in-flight annihilation events}
\label{GRASP_TOA}
\end{figure}

The grasps for antiprotons (black lines) and antideuterons (red lines) in terms of the energy at the top of the atmosphere are shown in Figure \ref{GRASP_TOA}. The solid lines are grasps for stopped events, while the dashed lines are grasps for in-flight annihilation events. Note that separate calculations show other antideuteron loss processes, such as Coulomb breakup or the Oppenheimer-Phillips effect, to be small ($<$ 10\%) \cite{Hailey2004}.

\subsection{Atomic X-ray}
\label{subsection: Atomic X-ray}

\begin{figure}[htbp]
\begin{center} 
\includegraphics*[width=7.5cm]{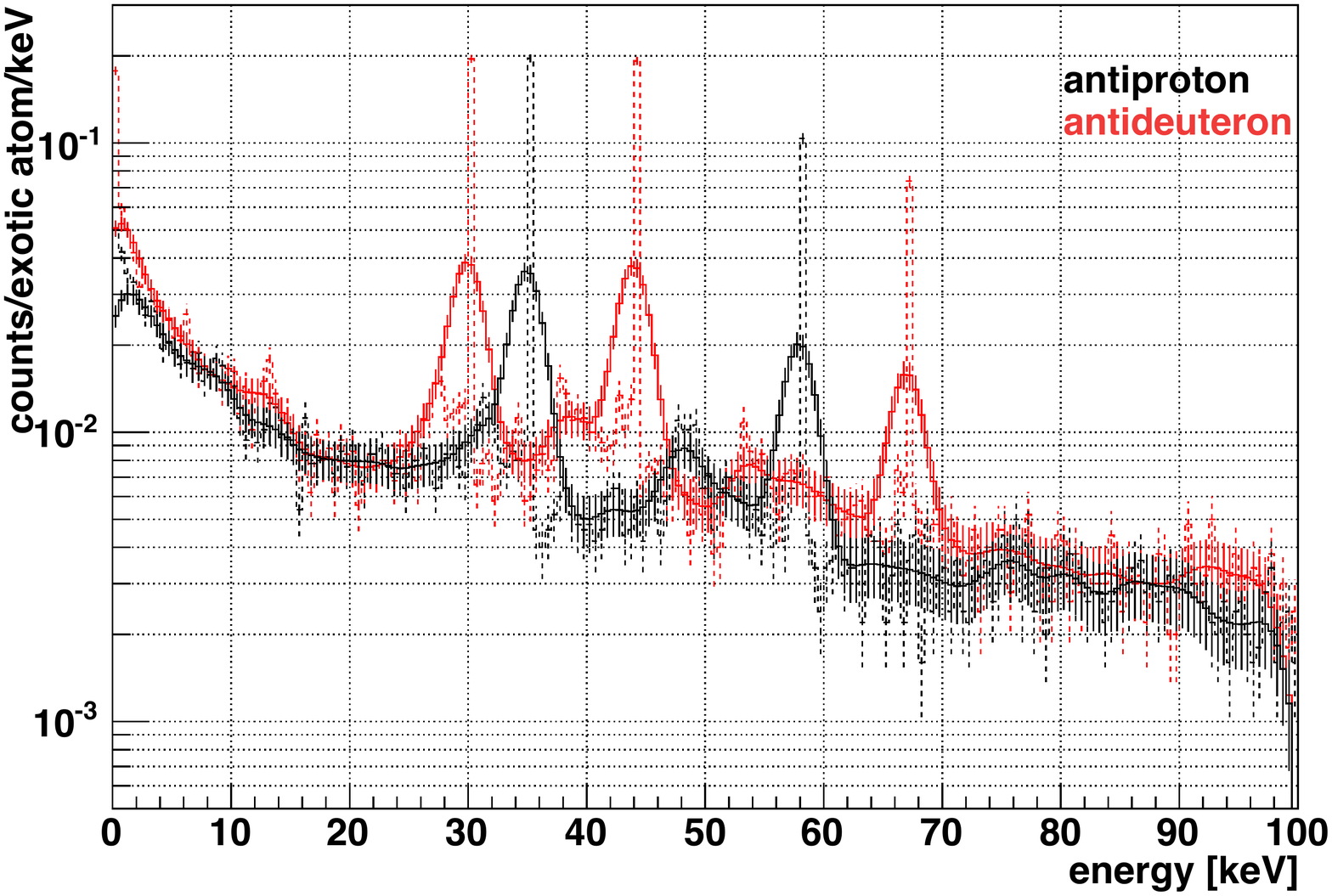}
\end{center}
\caption{Geant4 simulation results for energy spectra in the Si(Li) detector. The black and red lines are antiproton and antideuteron events, respectively. The dashed lines show energy spectra without detector response, while the solid lines taking into account the FWHM energy resolution ($\sim$ 3 keV) of the Si(Li) detector.}
\label{xray}
\end{figure}

As discussed above, the stopped antiparticles can form exotic atoms and emit atomic X-rays during the de-excitation. The energies of the atomic X-rays are uniquely determined by the components of the exotic atom as seen below. 
\begin{equation}
E_{X} = \left( zZ \right)^2 \frac{M^*}{m_e^*} R_H \left( \frac{1}{n^2_f} - \frac{1}{n^2_i} \right)  \tag{A}
\label{eq_xray}
\end{equation}
Here, $z$ and $Z$ are the charge of the antiparticle and target atom, $M^*$ and $m_e^*$ are the reduced masses of an antiparticle in the exotic atom and an electron in the target atom, $R_H$ is the Rydberg constant and $n_i$ and $n_f$ are the initial and final principal quantum number. The antideuteronic exotic atom formed with the Si target can emit 30 keV, 44 keV and 67 keV atomic X-rays, while 23 keV, 35 keV and 58 keV X-rays can be emitted for antiprotonic exotic atoms formed with the Si target. The X-ray yield, defined as the probability to emit atomic X-rays per exotic atom, for each atomic X-ray was estimated as $\sim$ 80\% with the simple cascade model as discussed in \cite{Aramaki2013}. Note that in addition to the antiparticles, atomic X-rays were manually generated in the simulation since the cascade model of the exotic atoms has not been implemented yet in the current Geant4 physics package. Figure \ref{xray} shows the simulation results for energy spectra in the Si(Li) detector, taking into account all the physics processes and interactions of particles and radiation passing through matter. The black and red lines are for antiproton and antideuteron events, respectively. The dashed lines show energy spectra without detector response, while the solid lines taking into account the FWHM energy resolution ($\sim$ 3 keV) of the Si(Li) detector. The continuous spectra are due to the EM shower developed by the annihilation products. The antideuteron selection cuts and efficiencies to atomic X-ray detection are shown in Table \ref{xray_efficiency}. The efficiencies to the detection of at least one antideuteronic atomic X-ray with these selection cuts were estimated as $\epsilon^{\bar{p}}_{X\geq 1} \sim$ 0.02 for antiprotons and $\epsilon^{\bar{d}}_{X\geq 1} \sim$ 0.12 for antideuteron events, as below. 

\begin{eqnarray*} 
\epsilon_{X\geq 1} = 1-(1-\epsilon_{30})(1-\epsilon_{44})(1-\epsilon_{67})
\end{eqnarray*}
Here, $\epsilon_{30}$, $\epsilon_{44}$ and $\epsilon_{67}$ are the efficiencies to antideuteronic atomic X-rays with the selection cuts. Note that atomic X-rays for antideuterons stopped in the detector frame and other materials are also distinguishable from those for antiprotons, since their atomic X-rays are different with each other, as seen in Eq (\ref{eq_xray}). The stopped position of the incoming antiparticle can be precisely determined by tracking and timing the annihilation products (charged pions) backwards in the detector layers and the TOF paddles.

\begin{table}[!h]
\caption{The select cuts and efficiencies to the antideuteronic atomic X-rays for antiproton and antideuteron events}
\begin{center}
\resizebox{8cm}{!}{%
\begin{tabular}{c|c|c|c}
Energy & selection cuts & $\epsilon^{\bar{p}}_{x}$ & $\epsilon^{\bar{d}}_{x}$ \\ \hline
30 keV & 29.5 $ \le E \le$ 30.5 keV & 9.5 $\times 10^{-3}$ & 3.8 $\times 10^{-2}$ \\ \hline
44 keV & 43 $ \le E \le$ 44.5 keV   & 8.0 $\times 10^{-3}$ & 5.4 $\times 10^{-2}$ \\ \hline
67 keV & 66 $ \le E \le$ 68 keV   & 6.7 $\times 10^{-3}$ & 3.0 $\times 10^{-2}$
\end{tabular}
}
\end{center}
\label{xray_efficiency}
\end{table}

\subsection{Nuclear Annihilation Products}
\label{subsection: Nuclear Annihilation Products}

\subsubsection{Intra-Nucear Cascade (INC) Model}
\label{subsubsection: Intra-Nucear Cascade (INC) Model}

\begin{figure}[htbp]
\begin{center} 
\includegraphics*[width=7.5cm]{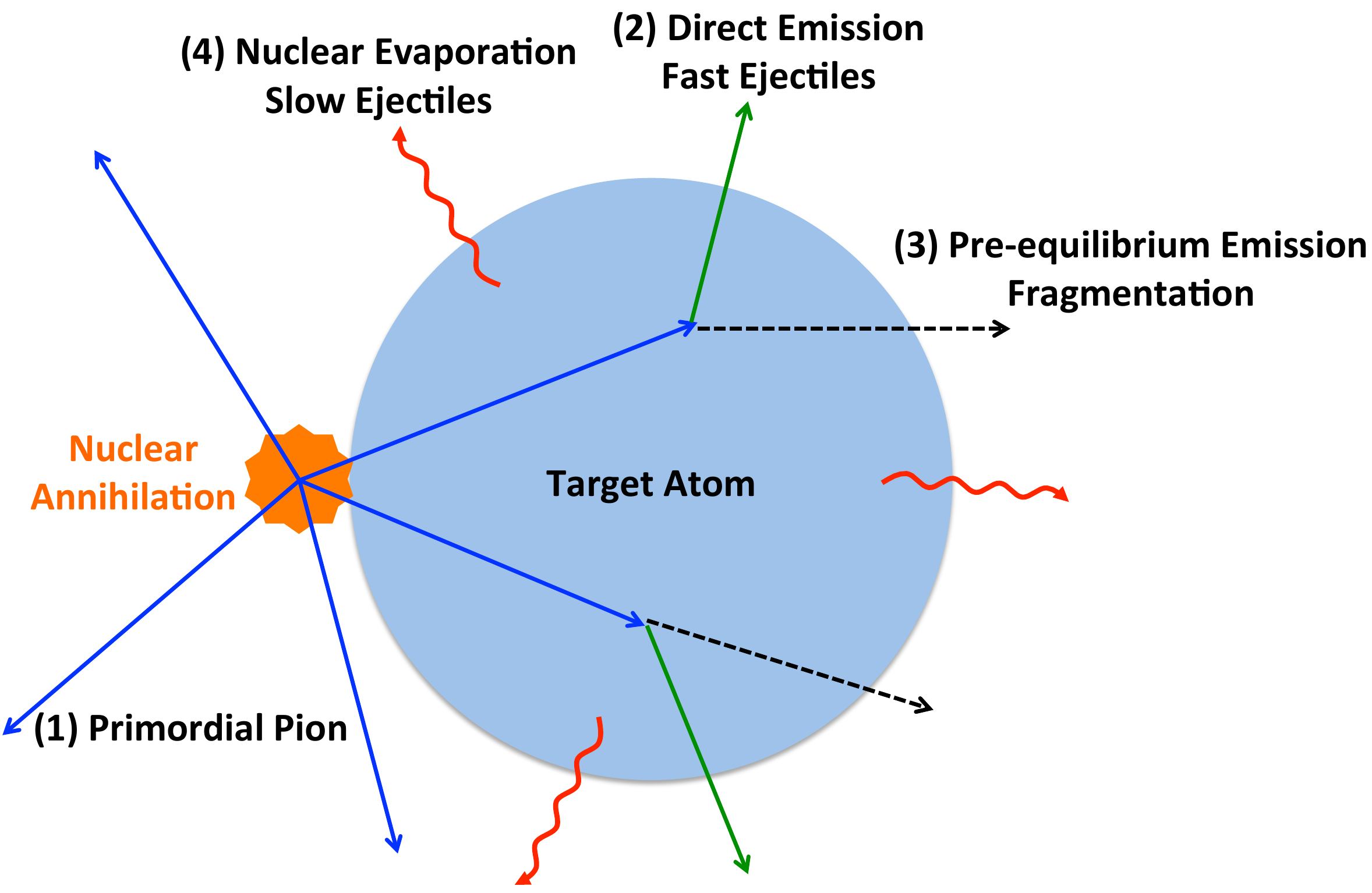}
\end{center}
\caption{Schematic of INC model. (1) The primordial pions are produced in the nuclear annihilation. (2) Some of the pions hit the nucleons in the nucleus with direct emission (fast ejectiles), followed by (3) the pre-equilibrium emission (fragmentation) and (4) the nuclear evaporation (slow ejectiles). }
\label{INCmodel}
\end{figure}

The interaction of antiprotons with nucleons (protons and neutrons) has been studied since the discovery of the antiproton in 1955 \cite{Chamberlain1955}. The Intra-Nuclear Cascade (INC) model has been developed to predict the particle multiplicity in the antiproton annihilation on nuclei \cite{Cugnon1989,Sudov1993}. The INC model is composed of four stages: (1) primordial pion production, (2) direct emission (fast ejectiles) from the primordial pion-nucleon interaction, (3) pre-equilibrium emission (multi-fragmentation) from excited nucleus, and (4) nuclear evaporation (slow ejectiles). The antiproton is first assumed to annihilate at the surface of a single nucleon of the nucleus and to produce the primordial pions. Some of these pions may escape from the nucleus, but the others cascade through the nucleons of the nucleus, knocking out fast nucleons as they go through. Due to this interaction, the nucleons in the nucleus are excited and their energy density is far from the thermal equilibrium state. Therefore, the nucleus can break into pieces with the emission of fragmented particles. The characteristic time of this process is $\sim$ 10$^{-22}$ s. Then the density distribution becomes more thermal and the remaining excitation energy will be removed by nuclear evaporation, which emits slow ejectiles. Figure \ref{INCmodel} shows an overview of the INC model processes. 

The INC model can also be applied to the antideuteron annihilation to estimate pion and proton multiplicities, resulting in two models that have been built based on how the two antinucleons in the antideuteron interact with nuclei \cite{Cugnon1992}. The first model (model A) assumes that the two antinucleons interact with the nucleons simultaneously, while the second model (model B) assumes that the antinucleons interact with nucleons separately. We will discuss how we can use the INC model to distinguish antideuterons from antiprotons in the following sections. The antideuteron selection cut and the efficiency were numerically evaluated.

\subsubsection{Pion Multiplicity}
\label{subsubsection: Pion Multiplicity}
The number of primordial charged and neutral pions in the INC model is estimated based on antiproton-nucleon annihilation. This has been well studied and much experimental data are available. The pion multiplicity and the standard deviation ($\sigma$) for the antiproton-nucleon annihilation is estimated as follows \cite{Cugnon1989,Cugnon1992}:

\begin{equation} 
\langle M^{p}_{\pi^{\pm,0}} \rangle = 2.65+3.65 \ln \sqrt{s} \tag{B}
\label{eq_mean}
\end{equation}
\begin{equation}
\frac{\sigma^2}{\langle M^{p}_{\pi^{\pm,0}} \rangle} = 0.174 \left( \sqrt{s} \right) ^{0.40} \tag{C}
\label{eq_sigma}
\end{equation}
Here, $\langle M^{p}_{\pi^{\pm,0}} \rangle$ is the average number of primordial charged and neutral pions and $\sqrt{s}$ is the center of mass energy in GeV. Eq (\ref{eq_mean}) and Eq (\ref{eq_sigma}) agree well with the experimental data \cite{Cugnon1989,Amsler1991,Amsler1998}. The average number of primordial charged and neutral pions $\langle M^{p}_{\pi^{\pm,0}} \rangle$ and the average number of primordial charged pions  $\langle M^{p}_{\pi^{\pm}} \rangle$ for the antiproton-nucleon annihilation at rest are 5.1 and 3.1, respectively. 

\begin{figure}[htbp]
\begin{center} 
\includegraphics*[width=7.5cm]{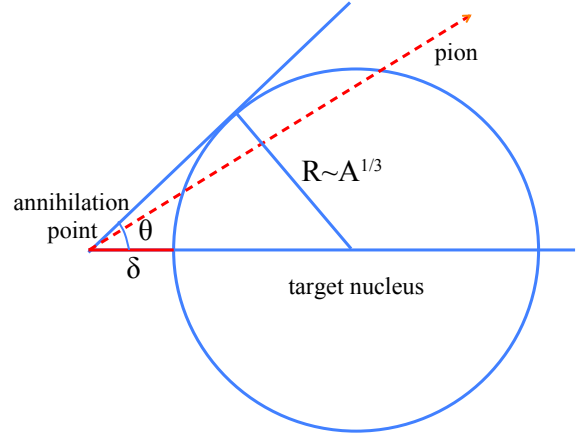}
\end{center}
\caption{Schematic of the interaction probability for primordial pions with the nucleons in the atom}
\label{INC_geometry}
\end{figure} 

Most of the primordial pions can escape from the nucleus, but some of them interact and cascade through the nucleons, producing emission of fast nucleons. The interaction probability can be geometrically calculated for each atom $A$, and the average number of the (final) charged pion multiplicity $\langle M_{\pi^{\pm}} \rangle$ can be simply estimated as follows \cite{Polster1995,Cugnon2001}:
 
\begin{eqnarray*}
\langle M_{\pi^{\pm}} \rangle &=& \langle M^{p}_{\pi^{\pm}} \rangle \cdot P(A) \\
P(A) &\sim& 1 -\frac{\Omega(A)}{4\pi}.
\end{eqnarray*}
\begin{eqnarray*}
\Omega(A) &=& 2\pi(1-cos\theta)  \\
&\sim& 2\pi \left( 1- \sqrt{1-\left( \frac{r_0A^{1/3}}{r_0A^{1/3}+\delta} \right)^2} \right)
\end{eqnarray*}
Here, $P(A)$ is the pion-nucleus interaction probability, $\Omega(A)$ is the solid angle that the pion can hit the nucleons in the atom, and $r_0$ and $\delta$ are the radius parameters, 1.2 fm and 1.6 fm, respectively \cite{Cugnon2001}. The INC model agrees well with the antiprotonic experimental data \cite{Polster1995}. Note that the pion multiplicity reduces by only 0.3 as the nuclei changes from $A$ = 10 to $A$ = 200. 

The primordial charged and neutral pion multiplicity for the antideuteron annihilation at rest on nuclei was estimated with Eq (\ref{eq_mean}) and Eq (\ref{eq_sigma}) by simply changing $\sqrt{s}$ = 4 GeV in model A and $\sqrt{s}$ = 2 GeV for each antinucleon in model B. The (final) charged pion multiplicity can be estimated with the pion-nucleus interaction probability as discussed above. The pion multiplicity and the efficiency to selection cuts for antiproton ($\epsilon^{\bar{p}}_{\pi}$) and antideuteron ($\epsilon^{\bar{d}}_{\pi}$) annihilations at rest on the Si target are shown in Table \ref{pion_multiplicity}. Here, we assume $\langle M_{\pi^{+}} \rangle = \langle M_{\pi^{-}} \rangle = \langle M_{\pi^{0}} \rangle$ and consider the average pion multiplicities between model A and B. The estimated pion multiplicity and the efficiency to selection cuts for in-flight annihilation events (annihilation with the kinetic energy T $\sim$ 0.1 GeV/n) were also shown in Table \ref{pion_multiplicity}.

\begin{table}[!h]
\caption{The pion multiplicity and the efficiency to selection cuts for antiproton and antideuteron annihilations at rest and in-flight (annihilation with the kinetic energy T $\sim$ 0.1 GeV/n) on the Si target.}
\begin{center}
\begin{tabular}{c|c|c||c|c}
 &  \multicolumn{2}{c||}{annihilation at rest} & \multicolumn{2}{c}{in-flight} \\ \hline
$\langle M_{\pi^{\pm}} \rangle$ & $\epsilon^{\bar{p}}_{\pi}$ & $\epsilon^{\bar{d}}_{\pi}$ & $\epsilon^{\bar{p}}_{\pi}$ & $\epsilon^{\bar{d}}_{\pi}$ \\ \hline
$\ge$ 3 & 0.53 & 0.93 & 0.57 & 0.93 \\ \hline
$\ge$ 4 & 0.25 & 0.80 & 0.58 & 0.82 \\ \hline
$\ge$ 5 & 0.070 & 0.62 & 0.092 & 0.64 \\ \hline
$\ge$ 6 & 0.093 & 0.41 & 0.019 & 0.44 \\ \hline
$\ge$ 7 & 0.00067 & 0.23 & 0.0024 & 0.26 \\ \hline
$\ge$ 8 & 0.000026 & 0.11 & 0.00016 & 0.13
\end{tabular}
\end{center}
\label{pion_multiplicity}
\end{table}

\subsubsection{Proton Multiplicity}
\label{subsubsection: Proton Multiplicity}

The INC model also predicts the proton and neutron production in the following processes: (1) direct emission from the interaction between the primordial pions and the nucleus, (2) pre-equilibrium emission (multi fragmentation) from excited nucleons, and (3) nuclear evaporation. The direct emission process dominates for protons with E $>$ 30 MeV and the energy spectrum for the proton multiplicity is estimated with the Maxwell-Boltzmann distribution as given below, based on fitting of the experimental data for antiproton annihilation at rest on nuclei \cite{Polster1995}:

\begin{eqnarray*}
\frac{dM}{dE} = \frac{2 \langle M_{p} \rangle}{\sqrt{\pi T^3}} \sqrt{E} \exp{\left( -\frac{E}{T} \right)}.
\end{eqnarray*}
Here, $\langle M \rangle$ is the average number of protons produced in the antiproton annihilation at rest on Si, $0.86 \pm 0.05$, $E$ is the energy of the proton, and $T$ is the parameter fitted to the data ($\sim$ 40 MeV). If the energy of the proton is too small, it will stop quickly and could not be tracked by the detector layers. Therefore, we set the lower cut of the proton energy as 60 MeV to guarantee passage through three or more Si(Li) layers. The proton multiplicity with $E \ge$ 60 MeV $\langle M^{60}_{p} \rangle$ is 0.37. Note that these protons are distinguishable from pions generated in the nuclear annihilation since, unlike pions, they are not relativistic and deposit much more energies in the detector layer. 

Unfortunately, there is no data available for the antideuteron annihilation at rest on Si. Cugnon et al.\cite{Cugnon1992}, however, estimated the nucleon (proton + neutron) multiplicity with $E \ge$ 60 MeV for the antideuteron annihilation at rest on the Mo (Molybdenum) target. The proton multiplicity for the Si target was simply estimated by scaling the multiplicity of the Mo target to the Si target by taking into account the pion-nucleus interaction probability as discussed above. We also assumed the relationship between the proton and neutron multiplicity as follows \cite{Polster1995}

\begin{eqnarray*}
\frac{\langle M_{n} \rangle/\langle M_{p} \rangle}{N/Z} \sim 2.
\end{eqnarray*}
Here, $N$ is the number of neutrons in the nucleus, and $Z$ is the atomic number. Considering the above, the proton multiplicity with $E \ge$ 60 MeV for the antideuteron annihilation on the Si atom was estimated as 2.35 for model A and 1.76 for model B. Table \ref{proton_multiplicity} shows the proton multiplicity with $E \ge 60 $ MeV and the efficiency to selection cuts for antiproton and antideuteron (average between model A and B) annihilations at rest on the Si atom. Here, the proton multiplicities for antiprotons and antideuterons were simply estimated as the Poisson distribution. The proton multiplicity and the efficiency to selection cuts for in-flight annihilation events (annihilation with the kinetic energy T $\sim$ 0.1 GeV/n) were also estimated simply by scaling the proton multiplicity based on the primordial pion multiplicity as seen in Eq (\ref{eq_mean}). 

\begin{table}[htdp]
\caption{The proton multiplicity with $E \ge 60 $ MeV and the efficiency to selection cuts for antiproton and antideuteron annihilations at rest and in-flight (annihilation with the kinetic energy T $\sim$ 0.1 GeV/n) on the Si atom.}
\begin{center}
\begin{tabular}{c|c|c||c|c}
 &  \multicolumn{2}{c||}{annihilation at rest} & \multicolumn{2}{c}{in-flight} \\ \hline
$\langle M_p^{60} \rangle$ & $\epsilon^{\bar{p}}_{p}$ & $\epsilon^{\bar{d}}_{p}$ & $\epsilon^{\bar{p}}_{p}$ & $\epsilon^{\bar{d}}_{p}$\\ \hline
$\ge$ 1 & 0.31 & 0.87 & 0.32 & 0.88 \\ \hline
$\ge$ 2 & 0.054 & 0.60 & 0.057 & 0.62 \\ \hline
$\ge$ 3 & 0.0064 & 0.34 & 0.0070 & 0.36 \\ \hline
$\ge$ 4 & 0.00058 & 0.16 & 0.00066 & 0.0.16 %\\ \hline
%$\ge$ 5 & 0.000043 & 0.062
\end{tabular}
\end{center}
\label{proton_multiplicity}
\end{table}%

\subsection{Depth Sensing and dE/dX Energy Deposit}
\label{subsection: Depth Sensing and dE/dX Energy Deposit}

The incoming particle can be tracked in the TOF paddles and Si(Li) detectors, as discussed above. The number of layers that the incoming particle can penetrate before stopping in the detector, provides the stopping range (called depth sensing) of the incoming particle. The stopping range for the antideuteron is approximately twice as large as that for the antiproton with the same velocity, as seen in the scaling law of the stopping range below \cite{Leo1987}.
\begin{eqnarray*}
R_{\bar{d}} \left( T_{\bar{d}} \right) = \frac{m_{\bar{d}}}{m_{\bar{p}}} \frac{z^2_{\bar{p}}}{z^2_{\bar{d}}} R_{\bar{p}} \left( T_{\bar{d}} \frac{m_{\bar{p}}}{m_{\bar{d}}} \right)
\label{range} 
\end{eqnarray*}
Here, $R_{\bar{d}}$ ($R_{\bar{p}}$) is the  stopping range of antideuterons (antiprotons), $m_{\bar{d}}$ ($m_{\bar{p}}$) is the mass of antideuterons (antiprotons), $z_{\bar{d}}$ ($z_{\bar{p}}$) is the charge of antideuterons (antiprotons) and $T_{\bar{d}}$ ($T_{\bar{p}}$) is the kinetic energy of antideuterons (antiprotons). Therefore, antideuterons can be distinguished from antiprotons by using the depth sensing and the TOF timing. Note that the stopped position of the incoming antiparticle can be precisely determined by tracking and timing the annihilation products (charged pions) backwards in the detector layers and the TOF paddles, as discussed above. 

A Geant4 simulation was conducted to evaluate the depth sensing for antideuteron selection cuts. Protons and deuterons were used to simply evaluate the depth sensing for antiprotons and antideuterons. The simulations were done for two different incoming angles, 0 deg and 45 deg, taking into account the timing and angular resolution of the TOF system, 0.5 ns and $\sim$ 5 deg. The tracking capability allows us to identify what material the incoming particle went through before stopping. 

\begin{figure}[htbp]
\begin{center} 
\includegraphics*[width=7.5cm]{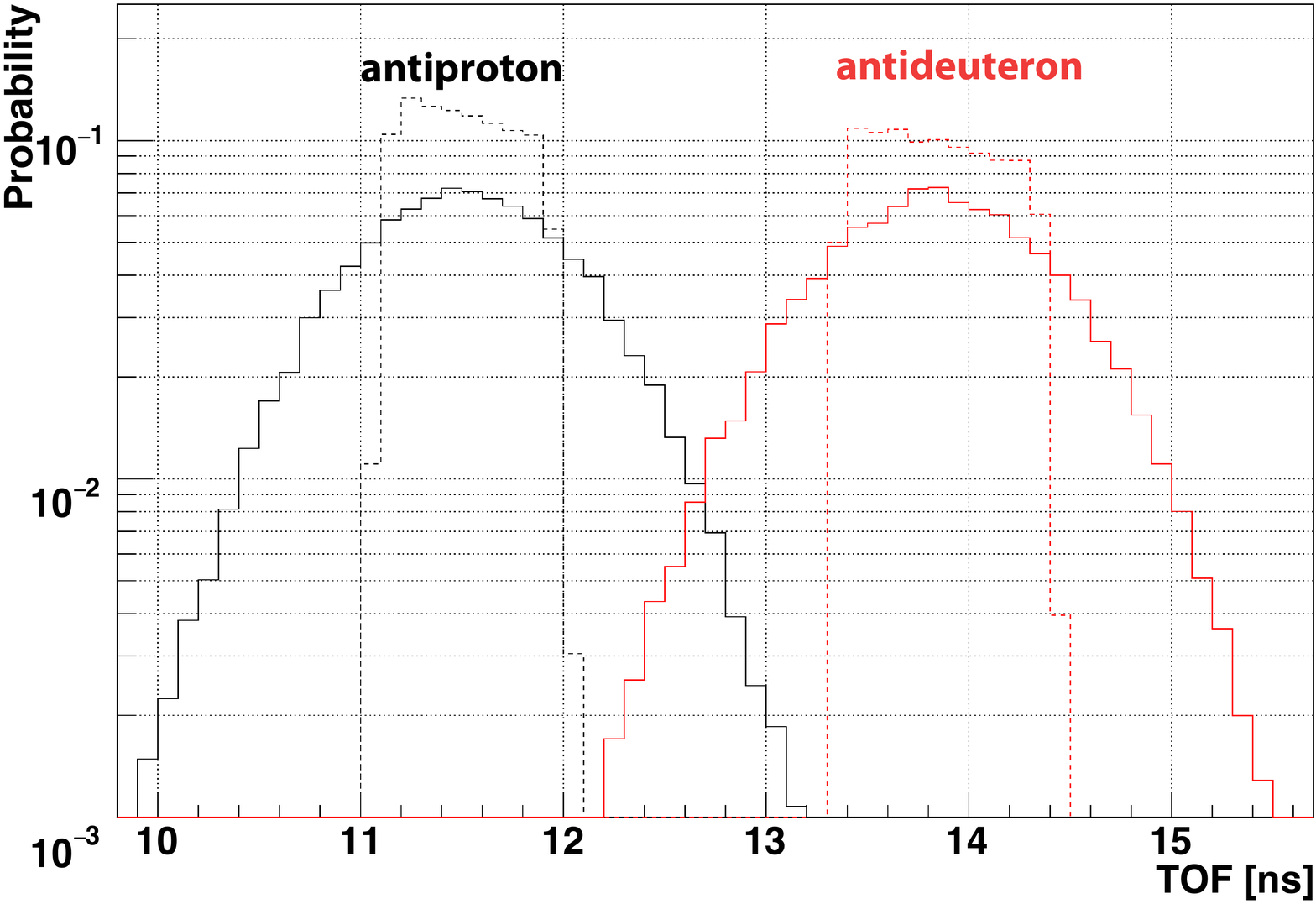}
\end{center}
\caption{The TOF timing for antiprotons (black lines) and antideuterons (red lines) with the incoming angle 0 deg and stopped in layer 3. The dashed lines show the TOF timing without detector response, while the solid lines taking into account the timing resolution (0.5 ns) of the TOF system}
\label{depth_0deg_3}
\end{figure}

\begin{figure}[htbp]
\begin{center} 
\includegraphics*[width=7.5cm]{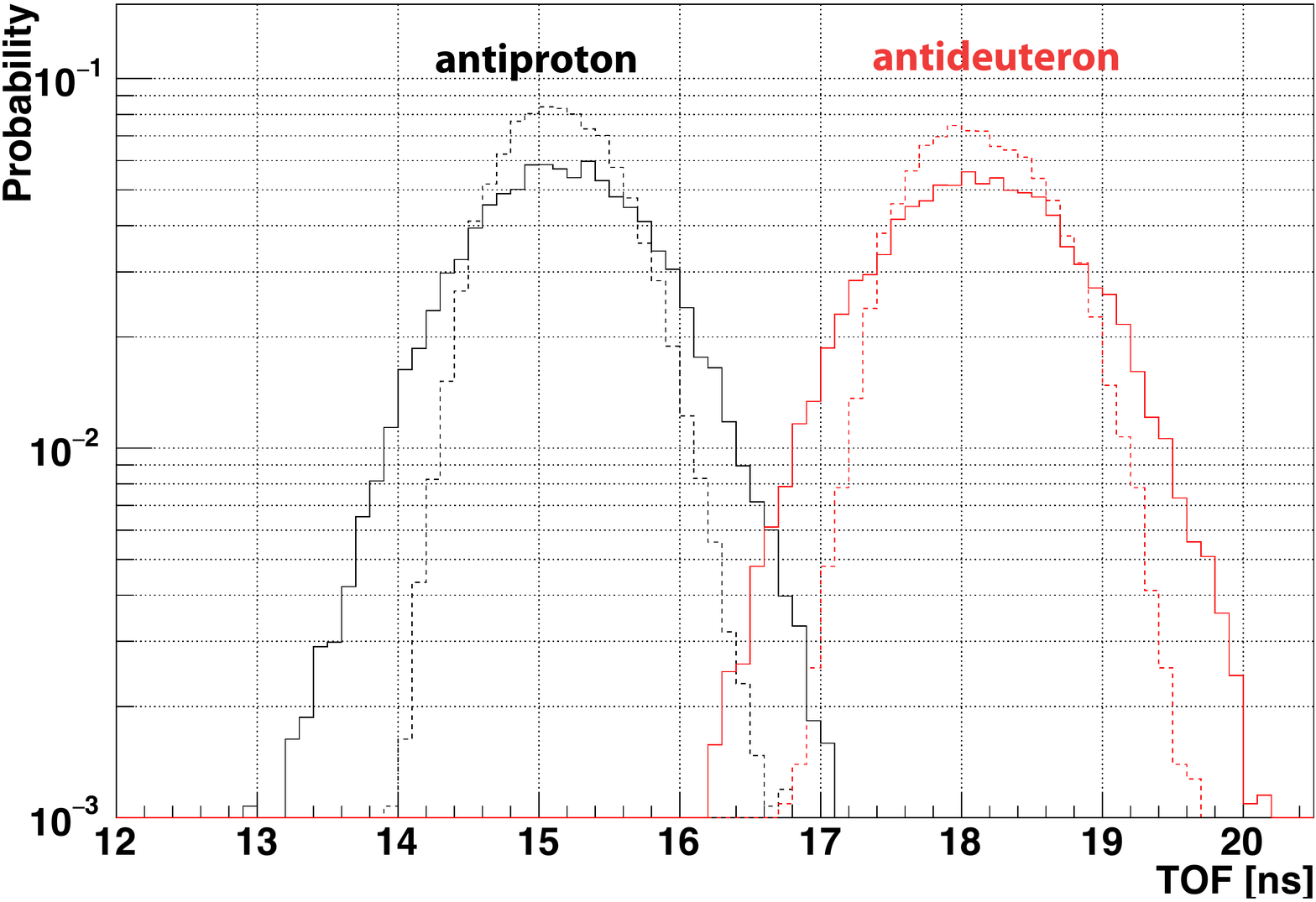}
\end{center}
\caption{The TOF timing for antiprotons (black lines) and antideuterons (red lines)  with the incoming angle 45 deg and stopped in layer 3. The dashed lines show the TOF timing without detector response, while the solid lines taking into account the timing resolution (0.5 ns) of the TOF system}
\label{depth_45deg_3}
\end{figure}

Figure \ref{depth_0deg_3} (\ref{depth_45deg_3}) shows the TOF timing for antiprotons (black lines) and antideuterons (red lines) with the incoming angle of 0 deg (45 deg) and stopped in layer 3. The dashed lines show the TOF timing without detector response, while the solid lines taking into account the timing resolution (0.5 ns) of the TOF system. Antideuterons stopped in the same layer as antiprotons show a larger TOF timing, as expected. The efficiency to the selection cut, $t_{TOF} \ge 13.3$ ns (17.6 ns) for antiparticles coming with 0 deg (45 deg) becomes $5.3 \times 10^{-4}$ ($6.1 \times 10^{-4}$) for antiprotons, while 0.83 (0.78) for antideuterons. 

Table \ref{depth_sensing_0deg_45deg} illustrates the antideuteron selection cuts and corresponding efficiencies for particles stopping in the different Si(Li) layers. The depth sensing provides excellent antideuteron identification from antiprotons, $\epsilon^{\bar{p}}_{depth} = 6.0 \times 10^{-4}$ and $\epsilon^{\bar{d}}_{depth} = 0.68 $ on average. Note that dE/dX energy deposit in the Si(Li) detector can be also used to distinguish antideuterons from antiprotons since antideuterons with the same TOF as antiprotons can have roughly twice as much kinetic energy as antiprotons, and deposit more energy in the Si(Li) detector. This will be evaluated with more detailed simulations and analyses, such as an event-by-event analysis and likelihood analysis, in the future. 

\begin{table}[htbp]
\caption{The antideuteron select cuts and the efficiency to the depth sensing for antiproton and antideuteron events. The incoming angles for antiparticles are 0 $\pm$ 5 deg and 45 $\pm$ 5 deg.}
\begin{center}
\resizebox{8cm}{!}{%
\begin{tabular}{c||c|c|c||c|c|c}
 &  \multicolumn{3}{c||}{0 deg} & \multicolumn{3}{c}{45 deg} \\ \hline
layer & cuts & $\epsilon^{\bar{p}}_{depth}$ & $\epsilon^{\bar{d}}_{depth} $ & cuts & $\epsilon^{\bar{p}}_{depth}$ & $\epsilon^{\bar{d}}_{depth} $  \\ \hline
1 & $\ge$ 16.7 ns & 5.9 $\times 10^{-4}$ & 0.66 & $\ge$ 24.2 ns & 7.2 $\times 10^{-4}$ & 0.82 \\ \hline
2 & $\ge$ 14.5 ns & 5.2 $\times 10^{-4}$ & 0.81 & $\ge$ 19.4 ns & 6.9 $\times 10^{-4}$ & 0.61 \\ \hline
3 & $\ge$ 13.3 ns & 5.3 $\times 10^{-4}$ & 0.83 & $\ge$ 17.6 ns & 6.1 $\times 10^{-4}$ & 0.78 \\ \hline
4 & $\ge$ 12.5 ns & 6.1 $\times 10^{-4}$ & 0.81 & $\ge$ 16.7 ns & 4.9 $\times 10^{-4}$ & 0.75 \\ \hline
5 & $\ge$ 12.0 ns & 5.4 $\times 10^{-4}$ & 0.74 & $\ge$ 15.6 ns & 4.6 $\times 10^{-4}$ & 0.79 \\ \hline
6 & $\ge$ 11.5 ns & 6.7 $\times 10^{-4}$ & 0.72 & $\ge$ 15.0 ns & 7.8 $\times 10^{-4}$ & 0.83 \\ \hline
7 & $\ge$ 11.1 ns & 7.7 $\times 10^{-4}$ & 0.70 & $\ge$ 14.6 ns & 6.3 $\times 10^{-4}$ & 0.77 \\ \hline
8 & $\ge$ 10.8 ns & 7.5 $\times 10^{-4}$ & 0.66 & $\ge$ 14.2 ns & 7.4 $\times 10^{-4}$ & 0.73 \\ \hline
9 & $\ge$ 10.7 ns & 4.7 $\times 10^{-4}$ & 0.51 & $\ge$ 13.8 ns & 7.5 $\times 10^{-4}$ & 0.72 \\ \hline
10 & $\ge$ 10.4 ns & 5.2 $\times 10^{-4}$ & 0.55 & $\ge$ 13.7 ns & 1.3 $\times 10^{-4}$ & 0.55 \\ \hline
ave & - & 6.0$\times 10^{-4}$ & 0.70 & - & 6.0 $\times 10^{-4}$ & 0.66 \\ 
\end{tabular}
}
\end{center}
\label{depth_sensing_0deg_45deg}
\end{table}

\subsection{Confidence Level and Sensitivity Calcuration}
\label{subsection: Sensitivity and Confidence Level}

The major background events in the GAPS antideuteron search are antiprotons that pass all the antideuteron selection cuts, as well as the secondary antideuterons produced by cosmic-ray interactions. The antiproton and the secondary antideuteron background events can be estimated as follows.

\begin{eqnarray*}
n^{bkg}_{\bar{p}} &=& \int F_{\bar{p}}(E) \Gamma_{\bar{p}}(E) T \epsilon^{\bar{p}}_{g} \epsilon_{T} dE \cdot \prod_{i} \epsilon^{\bar{p}}_{i}
\end{eqnarray*}

\begin{eqnarray*}
n^{bkg}_{\bar{d}} &=& \int F^{sec}_{\bar{d}}(E) \Gamma_{\bar{d}}(E) T \epsilon^{\bar{d}}_{g} \epsilon_{T} dE \cdot \prod_{i} \epsilon^{\bar{p}}_{i} 
\end{eqnarray*}
Here, $F_{\bar{p}}$ and $F^{sec}_{\bar{d}}$ are the antiproton and secondary antideuteron fluxes at the top of atmosphere, $\Gamma_{\bar{p}}$ and $\Gamma_{\bar{d}}$ are the grasps for antiprotons and antideuterons, $\epsilon^{\bar{d}}_{g}$ is the geomagnetic cutoff parameter for antideuterons, $T$ is the flight time, $\epsilon_{T}$ is the ratio of the observation time to the flight time ($\sim 1.0$), $\epsilon^{\bar{d}}_{i}$ is the efficiency to each cut and $i$ is the cut type (atomic X-ray, pion/proton multiplicity and depth sensing). Note that GAPS is proposed to fly from Antarctica, which is the best location with the lowest geomagnetic cutoff, $\epsilon^{\bar{p}}_{g} \sim 0.45$ for antiprotons and $\epsilon^{\bar{d}}_{g} \sim 0.75$ for antideuterons at $\sim$ 0.2 GeV/$n$ (evaluated with the Geant4-based PLANETOCOSMICS framework \cite{Desorgher2006,Doetinchem2013}). The sensitivity to $n$ antideuteron detections with the selection cuts, $S^n_{\bar{d}}$, and the corresponding confidence level (CL) to identify antideuterons from background events can be estimated as follows. 

\begin{eqnarray*}
S^n_{\bar{d}} &=& \frac{1}{\frac{1}{n}\int  \Gamma_{\bar{d}} (E) T \epsilon^{\bar{d}}_{g} \epsilon_{T} dE \cdot \prod_{i} \epsilon^{\bar{d}}_{i}} \\
CL &=& 1-P(N \ge n, \lambda = n^{bkg}_{\bar{p}}+n^{bkg}_{\bar{d}}) \\
&& P(N,\lambda) = \frac{e^{-\lambda}\lambda^N}{N!}
\end{eqnarray*}
Here, $P(N,\lambda)$ is the Poisson distribution with the observation value $N$ and the mean $\lambda$. The selection cuts for stopped events and in-flight annihilation events are different since in-flight annihilation events do not form exotic atoms. The total antideuteron sensitivity with $\sim$ 99\% CL, $S^{total}_{\bar{d}}$ was estimated as below. 

\begin{equation}
S^{total}_{\bar{d}} = \left( \frac{1}{S^{1}_{\bar{d},stop}} + \frac{1}{S^{2}_{\bar{d},inf}} \right)^{-1} \tag{D}
\label{1/S}
\end{equation}
Here, the antideuteron selection cuts for stopped events are ''depth sensing $+ \pi_{\ge 5} + \left( X_{\ge 1} \mbox{ or } p_{\ge 2} \mbox{ or } \pi_{\ge 2} \right)$'' for three LDB flights (LDB, 105 days), while they are ''$\pi_{\ge 6} + \left(p_{\ge 2} \mbox{ or } \pi_{\ge 2} \right)$'' for in-flight annihilation events. Two or more antideuteron detection is required for in-flight annihilation events to obtain $\sim$ 99\% CL. The corresponding antideuteron sensitivity and the number of background events are 2.0 $\times 10^{-6}$ [m$^{-2}$s$^{-1}$sr$^{-1}$(GeV/$n$)$^{-1}$] and $\sim$ 0.01 events, respectively. Note that the number of background events increases to 0.014 if atmospherically produced antiprotons and antideuterons are taken into account \cite{Duperray2005}. Atmospheric antiproton and antideuteron production and transport through the atmosphere and geomagnetic field will be studied more in the future.
 
The GAPS antideuteron sensitivity as well as antideuteron fluxes at the top of the atmosphere for different dark matter models are shown in Figure \ref{dbar_sensitivity}. Here, GAPS is sensitive to neutralino and gravitino LSPs in SUSY theories, as well as LZP in extra dimension theories, discussed in Section \ref{subsection: Antideuterons for Dark Matter Search}. Moreover, the GAPS antideuteron sensitivity is two orders of magnitude better than the current upper limit on the antideuteron flux obtained by the BESS experiment \cite{Fuke2005}. 

The AMS-02 antideuteron sensitivity (five years of observation time) was estimated for two different energy regions where two different detectors, the time of flight (0.2 GeV/$n$ $\le$ E $\le$ 0.8 GeV/$n$) and the ring imaging Cherenkov counter (2.2 GeV/$n$ $\le$ E $\le$ 4.2 GeV/$n$), are used for the velocity measurement \cite{Aramaki2015,Giovacchini2007}. Here, the geomagnetic cutoff at the International Space Station was taken into account for the sensitivity calculation. The antideuteron sensitivity in 2.2 GeV/$n$ $\le$ E $\le$ 3.6 GeV/$n$ is not as sensitive as the one in 3.6 GeV/$n$ $\le$ E $\le$ 4.2 GeV/$n$, due to the difference of the antideuteron identification capability. These two regions were not combined using Eq (\ref{1/S}). This was done to maintain consistency with the AMS-02 sensitivity calculations of \cite{Doetinchem2014,Mognet2014,Fuke2014,Aramaki2015}. All the sensitivities were estimated for $\sim$ 99\% CL to identify primary antideuterons from other cosmic-ray particles. Note that the AMS-02 sensitivity calculation is optimistic since it is based on the superconducting magnet, rather than the permanent magnet used in the actual flight.

\section{Conclusion}
\label{section: Conclusion}

Cosmic-ray antideuterons can provide a clean dark matter signature since the primary antideuteron flux due to the dark matter co-annihilation and decay can be a few orders of magnitude larger than the secondary flux at low energy. GAPS can uniquely probe dark matter through low-energy antideuterons with distinctive detection methods, while complementing existing and planned other indirect dark matter searches as well as underground direct dark matter searches experiments and collider experiments. The simultaneous detection of atomic X-rays and charged particles from the decay of exotic atom as well as the timing and depth sensing of the incoming particle provides excellent particle identification capabilities to select antideuterons from other cosmic-ray particles. The GAPS antideuteron sensitivity was estimated to be 2.0 $\times 10^{-6}$ [m$^{-2}$s$^{-1}$sr$^{-1}$(GeV/$n$)$^{-1}$] for three LDB flights (105 days, $\sim$ 99\% CL), which is more than two orders of magnitude better than the current upper limit on the flux obtained by BESS, and also sensitive to various dark matter models. 

\section{Acknowledgments}
\label{section: Acknowledgments}
This work is supported in the US by NASA APRA Grants (NNX09AC13G, NNX09AC16G) and the UCLA Division of Physical Sciences and in Japan by MEXT grants KAKENHI (22340073, 26707015). K. Perez's work is supported by the National Science Foundation under Award No. 1202958.



%% References


\bibliographystyle{model1-num-names}

%\bibliographystyle{named} 
\bibliography{refs}



%\begin{thebibliography}{00}

%\bibitem{Mori02} 
%K. Mori, C. J. Hailey, E. A. Baltz, et al.,
%A novel antimatter detector based on x-ray deexcitation of exotic atoms,
%ApJ 566, 604, 2002 

%\end{thebibliography}

\end{document}